%% file: main.tex
\documentclass[sigconf]{acmart} 
\usepackage{CJKutf8}
\usepackage{float}
\usepackage{subfigure}
\AtBeginDocument{%
  \providecommand\BibTeX{{%
    \normalfont B\kern-0.5em{\scshape i\kern-0.25em b}\kern-0.8em\TeX}}}
\usepackage{tabularx}
\usepackage{bbding}
\usepackage{graphicx}
\usepackage{geometry}
\usepackage{subfigure}
\usepackage{amsmath}
\usepackage{makecell}
\usepackage{float}
\usepackage{color}
\usepackage{booktabs}
\usepackage{caption}
\usepackage{tabu}
\usepackage{hyperref}
\usepackage{xcolor} 
\usepackage[normalem]{ulem} 

 \copyrightyear{2025} 
\acmYear{2025} 
\setcopyright{acmlicensed}\acmConference[C\&C '25]{Creativity and Cognition}{June 23--25, 2025}{Virtual, United Kingdom}
 \acmBooktitle{Creativity and Cognition (C\&C '25), June 23--25, 2025, Virtual, United Kingdom}
 \acmDOI{10.1145/3698061.3726920}
 \acmISBN{979-8-4007-1289-0/2025/06}

\author{Suifang Zhou}
\email{sfzhou3@my.cityu.edu.hk}
\orcid{0000-0001-5103-580X}
\affiliation{%
\institution{City University of Hong Kong}
\city{Hong Kong}
\country{China}}

\author{Kexue Fu}
\email{kexuefu2-c@my.cityu.edu.hk}
\orcid{0000-0002-2929-2663}
\affiliation{%
\institution{City University of Hong Kong}
\city{Hong Kong}
\country{China}}

\author{Huanmin YI}
\email{huanminyi2-c@my.cityu.edu.hk}
\orcid{0009-0005-1684-1642}
\affiliation{%
\institution{City University of Hong Kong}
\city{Hong Kong}
\country{China}}

\author{RAY LC}
\authornote{Correspondences should be addressed to LC@raylc.org.}
\email{LC@raylc.org}
\orcid{0000-0001-7310-8790}
\affiliation{
\institution{City University of Hong Kong}
\city{Hong Kong}
\country{China}}

\definecolor{fkx}{rgb}{0.1, 0.2, 0.8}

\begin{document}
\title[Retrochat]{RetroChat: Designing for the Preservation of Past Digital Experiences}



\begin{teaserfigure}
    \centering
    \includegraphics[width=\textwidth]{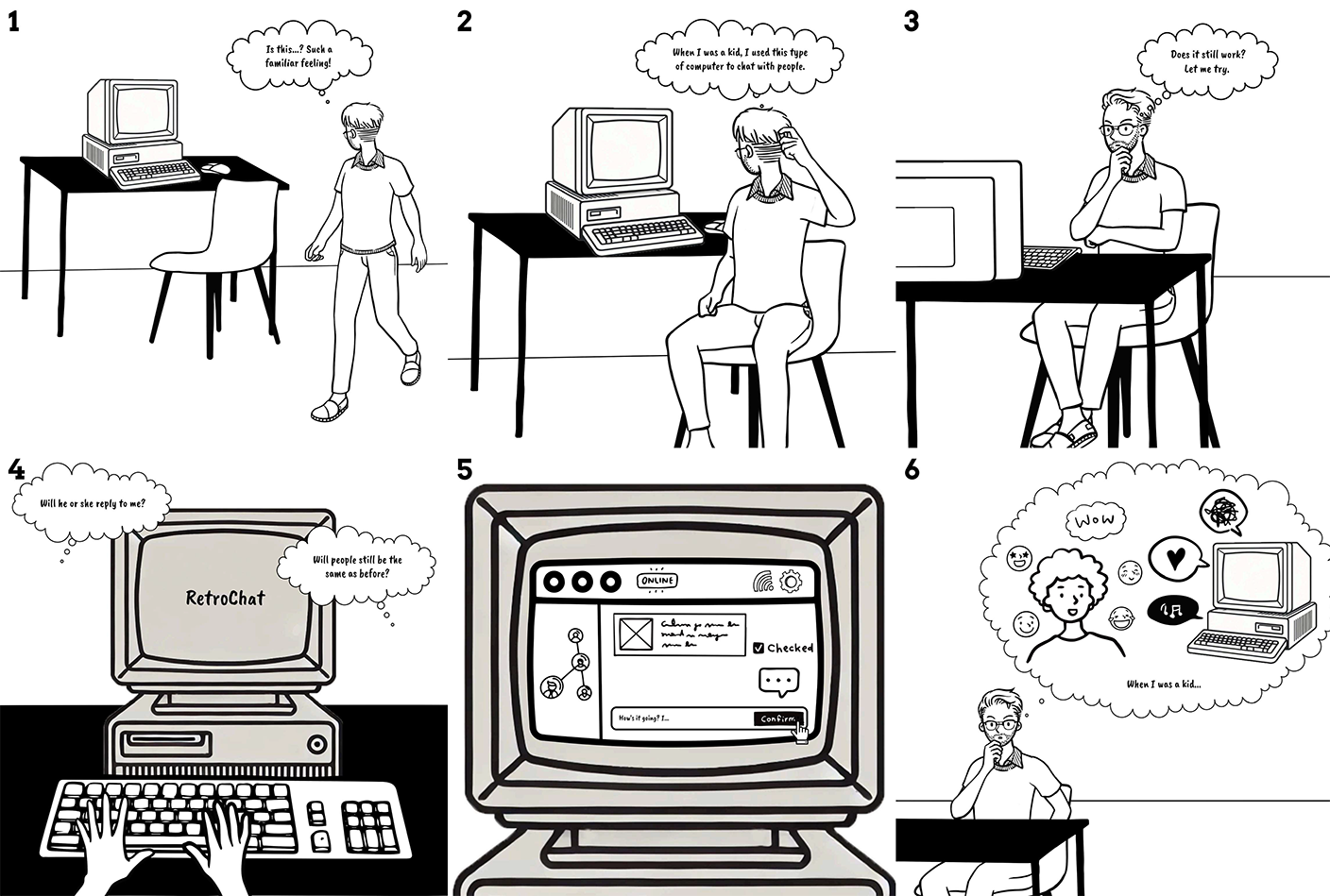}
\caption{Overview of RetroChat: Users interact with a chat agent that incorporates online expressions from Chinese social networks of the 2000s to 2010s, triggering memory flashbacks and in-depth reflections.}
    \label{fig:gamecover}
\end{teaserfigure}

\maketitle

\section*{abstract}

Rapid changes in social networks have transformed the way people express themselves, turning past neologisms, values, and mindsets embedded in these expressions into online heritage. How can we preserve these expressions as cultural heritage? Instead of traditional archiving methods for static material, we designed an interactive and experiential form of archiving for Chinese social networks. Using dialogue data from 2000-2010 on early Chinese social media, we developed a GPT-driven agent within a retro chat interface, emulating the language and expression style of the period for interaction. Results from a qualitative study with 18 participants show that the design captures the past chatting experience and evokes memory flashbacks and nostalgia feeling through conversation. Participants, particularly those familiar with the era, adapted their language to match the agent's chatting style. This study explores how the design of preservation methods for digital experiences can be informed by experiential representations supported by generative tools.

\ccsdesc[500]{Human-centered computing~Interactive systems and tools}

\keywords{Social Media, Cultural Heritage/History, Design Methods, Qualitative Methods}

\section{Introduction}\label{sec:Introduction}
\input{sections/01-Intro.tex}

\section{Related Work}\label{sec:Related work}

\input{sections/02-RelatedWork}

\section{Formative Study}\label{sec:Formative Study}
\input{sections/03-FormativeStudy}

\section{RetroChat}\label{sec:RetroChat}
\input{sections/04-RetroChat}

\section{Method}\label{sec:method}
\input{sections/05-Method}

\section{Results}\label{sec:Results}
\input{sections/06-Result}

\section{Discussion}\label{sec:Discussion}
\input{sections/07-Discussion}

\section{Conclusion}\label{sec:Conclusion}
\input{sections/08-Conclusion}


\bibliographystyle{ACM-Reference-Format}
\bibliography{references}

\end{document}

%% file: sections/01-Intro.tex
The introduction of the internet in China in 1994, followed by three decades of rapid computer mediated communication (CMC) development, has profoundly reshaped the nation’s cyber landscape \cite{herring2013introduction, yao2020computer}. Alongside this growth, efforts have been made to study cyber culture and  online communication. Researchers have observed that netizens invent unique way of expression, commonly referred to as "netspeaking" or "internet language." For instance, One common practice is to substitute Chinese characters with similar-sounding ones to change meanings and create internet slang, often forming distinct formulas that become widespread online \cite{wong2024digital}. However, as new online languages continually emerge and shift in response to trends \cite{zhang2024hǎo}, older expressions and linguistic styles gradually fade, with once-widespread phrases slowly disappearing from the online space \cite{goel2016social, varis2017internet, ziser2023rant, zanzotto2012language}.

This phenomenon of shifting and declining internet language has drawn increasing attention from online culture researchers \cite{he2008memes, knevzevic2023internet}. Scholars recognize that the disappearance of specific internet expressions and modes of communication signifies a loss of cultural elements\cite{acker2014death}. These communication forms encapsulate the behaviors, identities, and values of diverse internet subcultures, making them a vital part of digital heritage and essential for understanding cyber culture \cite{tan2024study, yan2021chinese, cannelli2022social, bennis2024cyberculture}.

For example, studies of internet culture and Chinese domestic cyberspace demonstrate that evolving online expressions are closely tied to changing in real-world social contexts. These expressions carry distinctive time imprints that reflect societal dynamics, collective mentalities, emotions, opinions, and ideologies, which are encapsulated in popular phrases and neologisms.\cite{sun2022toward, glushkova2018trends, sametouglu2024value, balakin2019internet}. This leads to a critical issue: as these way of expression and practices vanish, we risk losing significant qualitative insights into social expression within digital spaces \cite{tari2024leveraging, jeffrey2012new}.

Studies addressing the preservation of online expressions remain largely unexplored. Current efforts often rely on methods rooted in web archiving logic \cite{thomson2017preserving, acker2020social}. These approaches typically focus on capturing users, text, and corresponding metadata—such as timestamps and user profiles—as well as multimodal media, including pictures, videos, and raw text, using techniques like web crawlers and snapshots \cite{mahto2016dive, spiliotopoulos2012designing,pehlivan2021archiving}. While this approach facilitates the creation of large static data corpora for analyzing users and text \cite{finnemann2019web,huang2014data}, it overlooks the interactive and communicative nature of CMC.

Specifically, "texts"—which capture expressions and communication patterns—are continuously shaped by user interactions and engagement \cite{taprial2012understanding, lomborg2012researching}.
While the primary goal of traditional web archiving is to provide future researchers with resources for social network analysis, an interactive preservation approach would enable a first-person reexperience of the distinctiveness of internet language on social network platforms that have otherwise ceased to exist. This approach, through dialogue, potentially enables an exploration of how contemporary individuals reflect on and respond to these forgotten expressions, offering rich qualitative insights through interaction. This, in turn, helps researchers better understand the evolution of expressions in practice. Such an approach remains largely under-explored, presenting a critical design opportunity.

Enabling interactive forms that represent an era of online expression remains challenging due to technological limitations in capturing the subtle dynamics of natural language communication. However, recent advancements in large language models (LLMs) demonstrate impressive capabilities in mimicking human-like cognition, enabling researchers to simulate natural conversations between users and chat agents with desired characteristics through prompt engineering. \cite{sun2019fine,hou2024my}. AI systems like ChatGPT can create relatable characters with distinct personas that facilitate social network-style interpersonal dialogue, enabling meaningful and engaging conversation \cite{han2024teams, zhou2024eternagram, zhang2022storybuddy, zeng_ronaldos_2025}. These advancements present promising new opportunities for preserving dynamic forms of online communication. We proposes the following research questions:

\begin{itemize}
\item \textbf{RQ1:} How do we design an interactive form that captures the past online expression of the Chinese social media by utilizing LLM-based chat agents?
\item \textbf{RQ2:} How do users interact with LLM-driven chat agents that emulate the online expressive style of past Chinese SNS platforms?

\item \textbf{RQ3:} What are the challenges and opportunities of using interactive forms to preserve online expression in social media?
\end{itemize}

To address the research question, our study begins with a formative investigation to identify the characteristics of diminished netspeak and online expressions in the Chinese internet space while also gaining a deeper understanding of the context in which these interactions occurred. We conducted interviews with individuals who witnessed and engaged with the internet during its early stages in China, focusing on their experiences with social media and online communication over time. Our key findings indicate that the pre-2010 Chinese internet had a unique atmosphere and style of expression, distinctly different from today's online chatting experiences. We identified the preferred chat systems and platforms where online expression was commonly practiced during that era, as well as refined the chatbot persona to align with that time period. 

Building on insights from our formative study, we developed a design flow for prompt engineering and created RetroChat—a chat agent powered by GPT-4, deployed within a 2008 legacy version of Windows Messenger \footnote{https://escargot.chat/download/msn/} (MSN 8.1) running on the third-party service Escargot \footnote{https://escargot.chat/}. We asked participants to engage in free-form conversations with RetroChat and recorded their chat history. Additionally, we conducted retrospective interviews to gain insights into how participants interacted with the chatbot and how they perceived their experience after the task. We then applied thematic analysis to examine the data.

Our results indicate that participants re-engaged with the past by adapting their language and communication styles, naturally immersing themselves in nostalgic expressions. Moreover, interactions with RetroChat revealed that past online experiences could evoke digital living memories, extending beyond digital spaces into real-life recollections. The contribution of this study lies in both the design methodology and the exploration of AI-generated conversational systems as an innovative tool for enabling real-time observation of user engagement with legacy digital contexts. This resonates with the idea that digital preservation should embrace an interactive process—one that allows individuals to engage with memories by re-experiencing their own past interactions and exploring those of others.


%% file: sections/02-RelatedWork.tex
\subsection{Preserving for Social Media}
Billions of people worldwide engage with social media platforms like Twitter and WeChat to discover, discuss, and disseminate information. These platforms have transformed individuals from passive content consumers into active content creators, making social media an indispensable data source for researchers across the social sciences and other disciplines. Consequently, the preservation of social media content has garnered significant attention.

Current efforts to preserve social media predominantly focus on user-generated content and rely on archiving techniques borrowed from traditional web archiving \cite{brugger2011web}. This approach is advocated with the purpose of "archiving for future researchers," aiming to stabilize and capture constantly evolving information \cite{lomborg2012researching}. It is actualized through methods such as web scraping, snapshotting of social media content, and the construction of archiving systems for posterity \cite{pehlivan2021archiving, huber2011social, huang2014data}. Services like the Internet Archive's Wayback Machine \footnote{https://web.archive.org/} have extended their scope to include certain social media platforms, archiving publicly accessible content at specific points in time \cite{agarwal2022way}.

Static archiving methods are inherently limited in capturing the dynamic and interactive qualities of social media platforms that stem from real-time user interactions \cite{ogden2022everything, lomborg2012researching}. By overlooking these interactive elements, current preservation practices risk losing the essential attributes that characterize social media. Elements such as fleeting conversations, user-specific expressions, and evolving narratives are crucial for comprehending the social and cultural phenomena embedded within these digital environments.

An alternative approach to preserving social media content focuses on its media format—the social network platforms themselves, which are primarily software-driven and web-based. This perspective aligns with challenges in software preservation, such as those encountered in video game archiving and the conservation of digital interactive art \cite{di2017software, johansson2023video}. These fields extend beyond merely storing static artifacts; they aim to preserve the interactive behaviors, user experiences, and contextual engagement inherent in the original medium.

This means that effective preservation methodologies must go beyond capturing content alone; they must also account for the dynamic nature of user interactions and evolving conversations. This involves recording conversation threads, user engagements, and the contextual information embedded within dialogues. Techniques from game preservation and emulation environments provide potential solutions \cite{balcerzak2023real}. For instance, dynamic crawling allows for the continuous capture of changing content, while context-aware archiving ensures that relationships, metadata, and interlinked interactions are preserved to maintain contextual meaning. Additionally, emulation environments recreate the original platform experience by simulating its interface, functionalities, and user interactions \cite{carta2017metadata, pinchbeck_emulation_2009}. By integrating these approaches, social media preservation can more accurately reflect the fluid and participatory nature of digital communication.

\subsection{Chinese Online Expression and Internet Language as online heritage}
\begin{figure*}[h]
        \centering
        \includegraphics[width=0.99\linewidth]{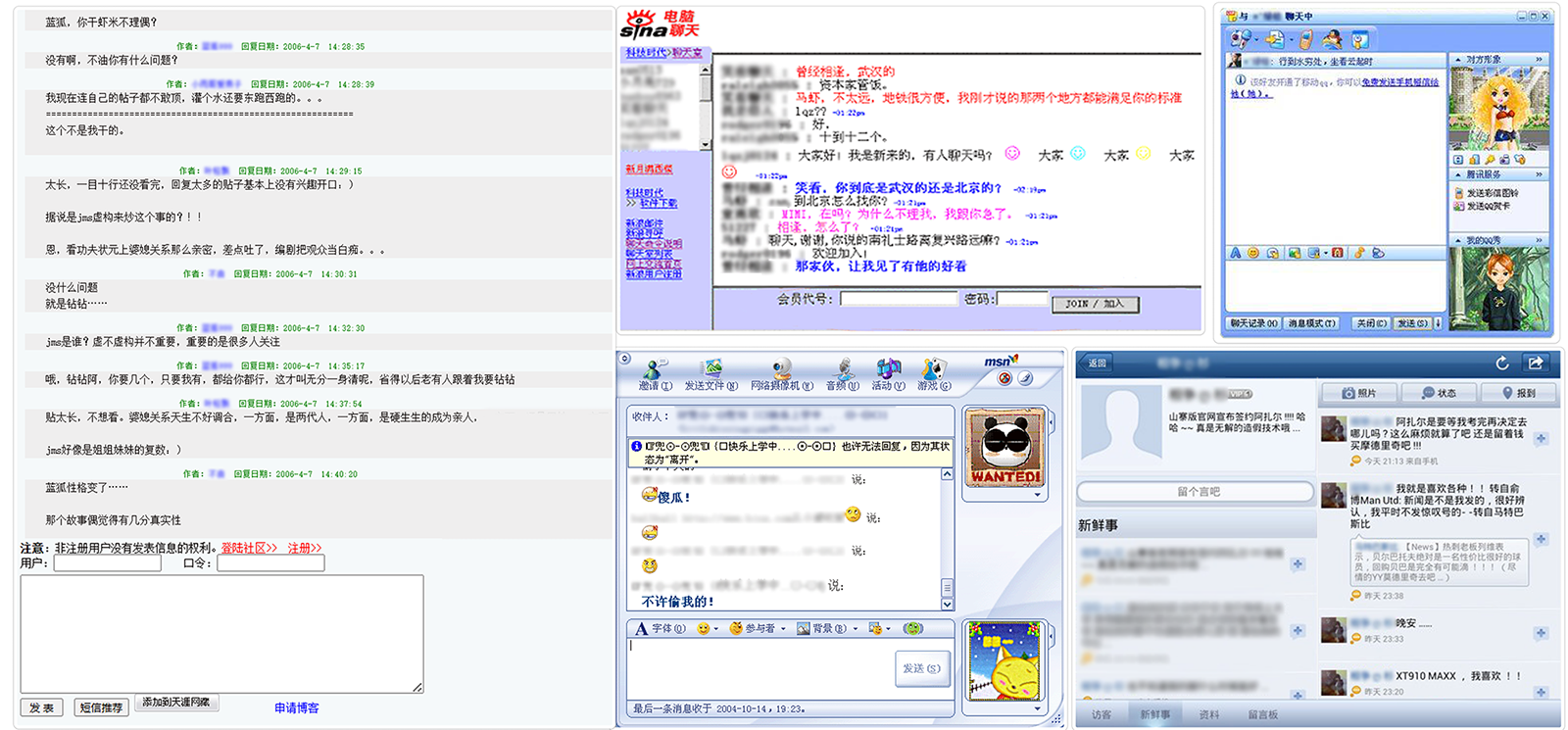}
        \caption{The figure presents user interfaces from early Chinese social networking platforms. On the left is a screenshot of the Tianya forum; the top middle displays the web-based Sina chatroom; the top right shows the 2005 version of QQ; the bottom middle features MSN Messenger 7.5; and the bottom right depicts the chat interface of Xiaonei on phone.}
        \label{fig:retro ui}
        \Description{Screenshots of early Chinese social platforms: Tianya forum, Sina chatroom, 2005 QQ, MSN Messenger 7.5, and Xiaonei mobile interface.}
    \end{figure*}

By December 2020, China had reached 989 million Internet users—a 43.7\% increase over the previous five years—making up nearly one-fifth of the global online population and establishing the world’s largest digital community \cite{xiao20222020}. This rapid expansion can be traced to the adoption of personal computers and the Internet in the 1990s, which saw widespread use by the early 2000s \cite{qiu2009working}. What emerges is a vibrant, multifaceted, and sometimes chaotic online culture that continues to evolve \cite{yang2003internet,lei2024playful}. This expansion has been closely intertwined with the proliferation of social networking technologies \cite{qiu2009working}, including a wide range of platforms and software applications (Figure ~\ref{fig:retro ui}). Collectively, these developments have significantly transformed modes of communication, interaction, and self-expression among internet users in China \cite{craig2021wanghong}. Today, Computer-Mediated Communication plays an essential role in daily life, serving as both a major source of information and a primary mode of personal interaction \cite{ shanmugasundaram2023impact}.

\setlength{\intextsep}{20pt} 

Research in sociolinguistics has highlighted how social factors contribute to the emergence of neologisms within Chinese online communities. For instance, lexical variations in "netspeak," often characterized by the homophonic substitution of conventional language forms, have been observed in Chinese social media. These linguistic innovations are frequently used to express heightened awareness of social issues \cite{he2008memes,chen2022internet}. A notable example is the expression \begin{CJK*}{UTF8}{gbsn}蒜你狠\end{CJK*} (suàn nǐhěn), which is a homophone of the conventional phrase \begin{CJK*}{UTF8}{gbsn}算你狠\end{CJK*} (suàn nǐhěn, “Fine, you win”) \cite{chen2022internet}. The netspeak variant emerged during a period of consumer anxiety in China, when rising prices of agricultural products, especially garlic (\begin{CJK*}{UTF8}{gbsn}蒜\end{CJK*}, suàn), sparked public awareness about inflation.

Further studies have demonstrated that social contexts often motivate the development of specific linguistic forms, with new meanings and qualities of expression becoming enregistered through shifts in both semantic and syntactic functions \cite{chen2020suicided}. These distinctive patterns are widely embraced by netizens and propagate in a memetic manner. One notable example is the phrase \begin{CJK*}{UTF8}{gbsn}被就业\end{CJK*} (bèi jiùyè, meaning "being employed"), which gained traction following reports that some colleges were unknowingly assigning graduates fake employment contracts to boost post-graduation employment statistics \cite{yao2013ironical}. This expression thus serves to satirize educational institutions as monetized, profit-driven entities \cite{tian2024ironic,guo2018playfulness}.

The prefix "bei" (\begin{CJK*}{UTF8}{gbsn}被\end{CJK*} ) has since evolved into a linguistic formula of "bei+XX" for expressing passive or involuntary actions, as seen in phrases such as \begin{CJK*}{UTF8}{gbsn}被自愿\end{CJK*} (bèi zìyuàn, "being taken as a volunteer"), and \begin{CJK*}{UTF8}{gbsn}被开心\end{CJK*} (bèi kāixīn, "being made to feel happy"). The insertion of the word "bei" (\begin{CJK*}{UTF8}{gbsn}被\end{CJK*}) before most content words, including nouns and adjectives, may initially appear grammatically unconventional \cite{huang2014new}. It nonetheless conveys a sense of resignation and frustration, often serving as a form of dark humor by subverting traditional linguistic norms to highlight these emotions\cite{guo2018playfulness,wang2021english}.

Research has highlighted connections between online language use and subcultural identity. The subculture known as 'FZL'—an abbreviation for \begin{CJK*}{UTF8}{gbsn}非主流\end{CJK*} (Fei Zhu Liu, or 'non-mainstream')—emerged among Chinese netzens in the early 2000s as a distinct form of digital self-expression \cite{zhou2009being}. This now-fading trend featured creatively altered, glitch-like Chinese characters, commonly called huoxingwen (\begin{CJK*}{UTF8}{gbsn}火星文\end{CJK*}, or 'Martian language') \cite{zitong2013cuteness}. 

This form of expression has garnered increasing attention from scholars—including those in documentary studies and sociology—who have highlighted its roots in social inequalities \cite{li2023dv, yuan2021web}. Many individuals who identify with the FZL subculture are second-generation young migrant workers who have endured harsh working conditions and societal alienation in urban China \cite{zhou2009being, liu2012negotiating}. In response, they have adopted niche aesthetics, allowing their online expressions to be interpreted as acts of resistance—a uniquely Chinese, working-class interpretation of punk-inspired, proactive differentiation against tangible discrimination and inequality \cite{tian2019allure, liu2014shamate, li2023dv}.

\subsection{AI for Natural Conversation Dialogue}

Anthropomorphism refers to the tendency to attribute human characteristics, such as emotions, awareness, or cognitive abilities, to non-human entities~\cite{epley2007seeing}. This concept has been increasingly integrated into chatbot design. Chatbots have evolved from rule-based systems like ELIZA~\cite{weizenbaum1966eliza} to deep learning-driven conversational agents capable of context-aware dialogue~\cite{zhao2023survey}. Modern chatbots, particularly those powered by large language models (LLMs), utilize pre-trained knowledge to generate human-like responses, enabling more fluid and personalized interactions~\cite{liu2024proactive,shumanov2021making,park2023generative}. Interactions with role-playing chatbots can influence users’ perceptions and behaviors. Recent findings highlight that chatbots improve user engagement by mimicking human conversational patterns, such as maintaining cohesive dialogue styles, transforming them from simple information retrieval tools into emotional companions~\cite{svikhnushina2022peace,skjuve2021my,yang2019understanding}. Prior research suggests that chatbots employing anthropomorphic language styles can enhance user compliance, indicating that role-playing AI has the potential to shape user behavior~\cite{adam2021ai}. 
Projects like "Shelly," a community-driven AI dedicated to horror storytelling, exemplify the development of AI companions designed for specific literary genres~\cite{yanardag2021shelley}. In the domain of role-playing chatbots, Replika—a companion chatbot designed to foster virtual romantic relationships—has been shown to enhance user satisfaction and involvement. Additionally, Valtolina and Marchionna~\cite{valtolina2021design} explored the role of Charlie, a chatbot designed for role-playing and providing emotional companionship to the elderly, demonstrating its effectiveness in alleviating loneliness. Another study demonstrated that engaging with a GPT-driven chatbot system within a game-like design framework can effectively assess attitude changes~\cite{zhou_eternagram_2024}. These advancements contribute to a greater sense of coherence in chatbot-driven conversations, making interactions more immersive and engaging.

Apart from the impact of communicating with personalized, human-like AI on behavior and perception, as well as its ability to enhance engagement and immersiveness \cite{yang_ai_2022}, users’ conversational behaviors were mainly influenced by chatbots’ communication styles~\cite{poivet2023influence}. Existing research primarily focuses on enhancing chatbots' ability to mimic human conversational styles~\cite{ait2023power} from a technical perspective and improving their role-playing capabilities~\cite{lu2024large}. However, fewer studies have examined how human language adapts and the conversational changes when interacting with chatbots. 
A phenomenon known as Language Style Matching (LSM) has been observed in human-human interactions ~\cite{niederhoffer2002linguistic}. The mimic conversational behavior can be explained by Communication Accommodation Theory (CAT) which suggests that humans tend to adapt to their conversation partners' tone, vocabulary choices, and even grammatical structures during interactions. This implies that when interacting with a chatbot that exhibits a specific language style, users may engage in linguistic mimicry. However, how users adapt their language when interacting with AI that take on specific roles remains largely unexplored.

Recent studies have highlighted the unique role of Generative AI(GenAI)'s potential in enhancing users’ ability to express themselves and recall personal experiences. A recent work~\cite{fu2024being} gained empirical data through a pre-designed iterative  collaborative workflow between AI and people when envision future scenarios of cultural heritage, and find that participants tend to reflect on their past experiences and recall personalized expressions during the interaction process. In another study, researchers conducted workshops to explore GenAI's ability to assist expression of local narratives of cultural heritage, the results highlighted GenAI's strengths in illuminating, amplifying, and reinterpreting personal narratives~\cite{he2024recall}.  These findings highlight the impact of interacting with GenAI on users' expression and personal memory recall, and suggest its potential as a means of intangible value of cultural heritage preservation.

%% file: sections/03-FormativeStudy.tex
Our attempted design focuses on enabling interactive dialogue within outdated social media platforms—a subject not extensively explored in existing research, particularly in the context of Chinese social network. To address this, we conducted a formative study using interviews to explore key design strategies for effectively building interactive preservation for past chatting experience.

First, it is essential to narrow down the definition of diminished netspeak and online expressions. Current studies on Chinese internet language often examine linguistic shifts, such as semantic variation, along with societal factors that influence the emergence or disappearance of certain online expressions. However, little effort has been made to specify the time range that would qualify these expressions as distinctive to a particular era.

From a methodological standpoint, the lack of a clear definition of online expressions raises issues of validity, specifically questioning what is being preserved through our design. From a design perspective, it directly influences how we utilize historical dialogue data for prompt engineering. These interconnected challenges highlight the importance of defining the temporal scope of online expressions to ensure clarity and relevance in both research and design methodologies.

Second, the interactive preservation of internet language necessitates consideration of the platforms where these conversational interaction occur. This raises the question: in what contexts are these languages used? Do users employ netspeak uniformly across cyberspace, or are such online expressions platform-specific? For example, instant messaging systems like MSN and QQ, bulletin board systems like MOP and Tianya, or blogging platforms like Sina Blog. We consider this aspect essential, as our effort to emulate interaction involves deploying chat agents within the appropriate platform and context.

\subsection{Participants} We located five participants for the formative study, all native Chinese speakers, recruited from the researchers' social network using Wechat Moments. They expressed interest in the topic and reported having experience and active engagement on social media since the early phase of Chinese social networks. Their familiarity with early Chinese social networks was assessed through a screening process. As part of this process, we provided a sample dialogue retrieved from early Chinese social networking platforms to evaluate their understanding and practical experience with internet language, including its contextual applications. Each participant was compensated with 50 Chinese Renminbi or the equivalent in Hong Kong dollars for their participation.

\begin{table*}[h!]
\centering
\resizebox{0.99\textwidth}{!}{%
\begin{tabular}{|p{2cm}|p{2cm}|p{2cm}|p{6cm}|p{3.5cm}|}
\hline
\textbf{Participants} & \textbf{Age} & \textbf{Gender} & \textbf{Most Used early SNS platform} & \textbf{First year using SNS} \\ \hline
\textbf{FSp1} & 35 & Female & QQ, MSN, Xiaonei, BBS, Chatrooms & 2002 \\ \hline
\textbf{FSp2} & 43 & Male & MSN, Yahoo! Chat, BBS & before 2000\\ \hline
\textbf{FSp3} & 34 & Female & QQ, Kaixin, Fetion & No later than 2005\\ \hline
\textbf{FSp4} & 41 & Male & QQ, MSN, BBS & No later than 2002\\ \hline
\textbf{FSp5} & 37 & Male & QQ, Web Chatrooms, Kaixin & 2002\\ \hline
\end{tabular}
}
\caption{Participants who took part in the formative interview}
\Description{Summary of 5 participants in the formative interview phase.}
\end{table*}

\subsection{Procedure} 

The interviews were conducted remotely via Tencent Voov Meeting, each lasting approximately 30 minutes. Participants were presented with user interface image sets of most used social network platforms beginning from the year 2000, based on data from the Statistical Report on China's Internet Development released by CNNIC \footnote{https://www.cnnic.com.cn/IDR/ReportDownloads/}. We included this platform demonstration to help participants recall their past experiences with online social platforms. After interacting with the system, they were asked to reflect on their memories and experiences, with particular emphasis on their practices of online expression and what made past chatting experiences unique or distinct from contemporary social interactions. We also presented dialogue samples from different years to gather their perspectives on the distinctiveness of online expressions over time.

\subsection{Key Findings} 

\subsubsection{KF1: Online Expressions Are Preferred in Close Circles or When Anonymous}

During our interviews, four participants shared their thoughts on internet slang usage and its connection to different platforms. FSp1 reflected, "I mostly use internet slang with my friends or when I am anonymous. I avoid using netspeak or "bad" internet slang on Xiaonei since most users there are my classmates and teachers, so I tend to be more formal." Similarly, FSp5 shared, "I mainly use that kind of language with friends and gaming buddies on QQ." FSp2 added, "BBS is where I remember seeing a lot of online expressions." FSp4 noted the same, "I don’t often use buzzwords now, but I used to see them all the time on BBS forums like Tianya and other discussion boards, where users were anonymous and real-life identities were unknown."

\subsubsection{KF2: Pre-2010 Online communication Carry a Distinct Mark of their time.}

We identified an intriguing trend when participants reflected on era-specific online expressions. All participants observed that today’s Chinese online expressions differ significantly from those of early netspeak. FSp1 and FSp3 both remarked, "At a glance, I can tell it’s not from today's internet." Similarly, FSp5 noted, "People nowadays no longer talk like that online." FSp4 added depth to this observation, commenting, "The language felt more upbeat back then, almost in a naïve way." FSp2 provided context, suggesting, "When netspeak first emerged, it was a trendy thing—using it conveyed a sense of, "Wow, you are surfing online too?" But now, we have gotten used to it."

When asked to highlight changes over the years, a recurring theme of pre-2010 Chinese online expressions emerged. FSp1 described this era as a "retro-style Chinese online climate," reminiscing about the vibrant activity on BBS forums and open chatrooms, which she attributed partly to "less awareness of cybersecurity back then." She described it as, "By security thing, I mean that people were less aware that chatting online could affect their real life. Now, we have things like 'RenRou' happening."  FSp4 expressed nostalgia for 2010, stating, "... Back then, people communicated more frankly through words. Now, many emotions and words have been replaced by GIFs and memes." Similarly, FSp5, during reviewing chat samples from 2005–2009, reflected, "People used more direct language, and some expressions would be considered intrusive or even vulgar by today's standards."

\subsubsection{KF3: How online identity is represented in cyber spaces is different.}

Two participants reflected on how online identity was represented in early digital spaces. When shown user interfaces from various chatting platforms, FSp1 remarked on avatars, saying, "It's so simple, almost like it was made with Microsoft Paint... This type of avatar that users were once so fond of no longer exists, unless someone intentionally adopts it for a retro style nowadays." He added, "I remember dozens of friends I met online using this avatar set to represent themselves." FSp3 shared a similar sentiment, stating, "I used the bum-head avatar for many years. It’s such an iconic design—it can be the first image that comes to mind when someone mentions retro Chinese social networks." She also highlighted the role of nick names in expressing identity: "I think nick names were another key part of early cyberspace identity. Coming up with a standout name was a big deal." She elaborated, "Since avatars often came in predefined sets with limited customization, people got creative with unique and quirky nick names to stand out."

\subsubsection{KF4: Distinctive Context and communication Device of Early Online Chatting}

When reflecting on the context of early online chatting, participants shared vivid memories of their experiences. FSp5 described a nostalgic scene, saying, "...because it was forbidden by my parents to go to the Wang Ba, now called web cafés, I often skipped class to game on chunky monitors." He added, "A big part of my online chatting happened on PCs. In QQ Fantasy I meet new friends and add them on QQ." FSp2 highlighted the functional distinctions between devices, stating, "Most my chatting for social purpose happened on desktops." He elaborated on the transitional period before the rise of phone-based social platforms, noting, "There was a time before phones became powerful. Communication on phones and computers served different purposes—phones were for messaging, while computers were for real socializing."

\subsubsection{KF5: Internet Slang are fading away rather than disappearing completely}

We found that outdated internet slang don’t entirely vanish from cyberspace but instead gradually fade from prominence. As FSp3 observed, "Some of those slangs and phrases are still in use, just not as popular as when they first emerged." FSp1 added, "If you hadn’t shown them, most would feel unfamiliar now… although some expressions, especially certain neologisms, have managed to survive over time." FSp2 highlighted a similar pattern during the early years of Chinese social networks, noting, "I remember when the SNS was just starting in China, there were more ways of using numbers to represent Chinese characters. Some of those expressions quickly fell out of use, but some are still around today."

\subsection{Design Strategies} 

Based on the key findings from the formative study phase, we identified six key design strategies to better support the creation of interactive systems for preserving the chatting experience, particularly in prompting engagement, chatbot personas, and the construction of the chatting environment:

\begin{itemize}
    \item \textbf{DS1}: Design prompts for chat agents aligned with identified time ranges to effectively represent the perceived past Chinese online chatting experience.

    \item \textbf{DS2}: Conversational data sourced from platforms with lower social pressure, such as BBS forums and instant chat systems, where natural internet expressions emerge organically among friends or netizens with fewer social constraints.
    
    \item \textbf{DS3}: Incorporate a graduated structure in prompt Internet Slang that reflects the frequency usage over time, mirroring the fading nature of Slang and formular way of expression online.
    
    \item \textbf{DS4}: The design of chat agents can incorporate personas and appearances that reflect the social characteristics of their respective time periods.
    
    \item \textbf{DS5}: The emulation of the chatting software need to be visually and functionally aligned with the specific time range.
    
    \item \textbf{DS6}: As nostalgia-driven factors influence how people recognize and interpret their experiences, the design of authenticity the chatting environment should be take into consideration.
\end{itemize}

%% file: sections/04-RetroChat.tex
\begin{figure*}[h]
        \centering
        \includegraphics[width=0.99\linewidth]{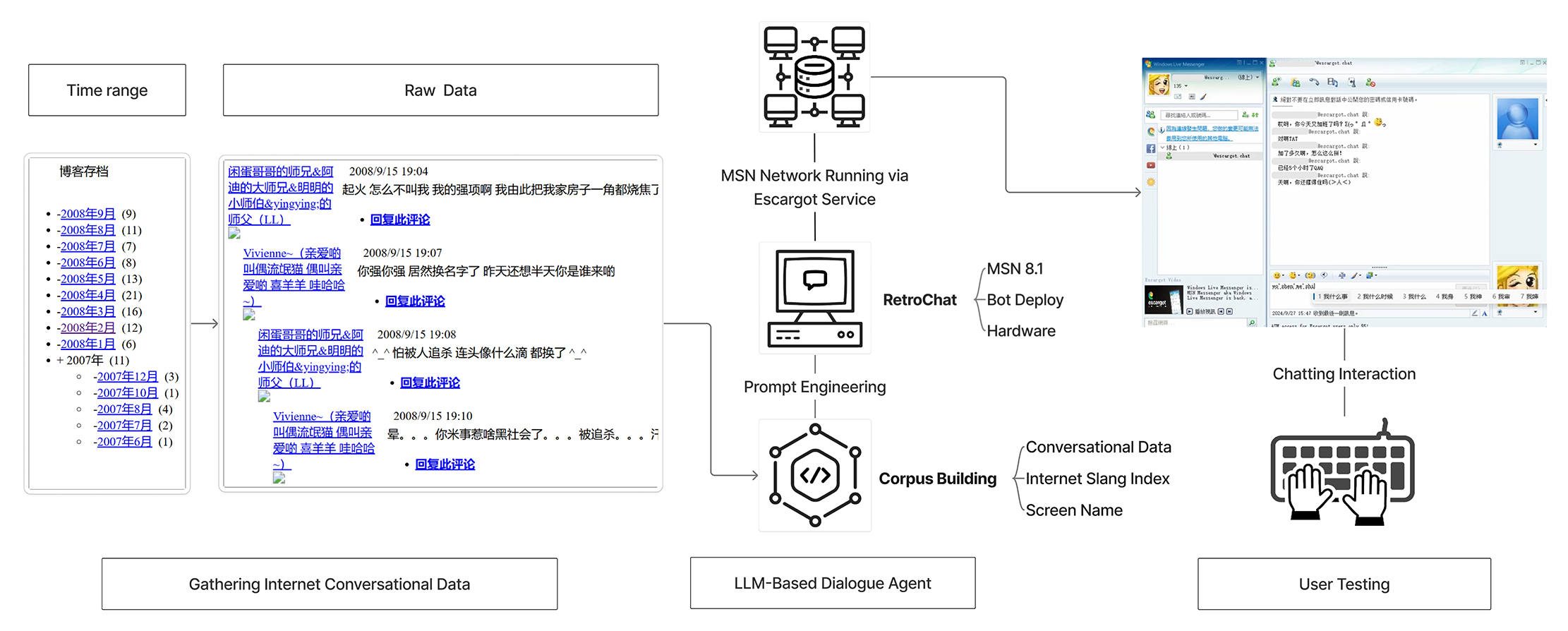}
        \caption{The diagram indicates the sequence of our design, starting from the initial gathering of raw data, followed by corpus building based on the dataset, prompt engineering, and finally deploying RetroChat through an MSN service to enable interaction.}
        \label{fig:Flow v0}
        \Description{Diagram illustrating the process from data collection to corpus building for prompt engineering, followed by the deployment of RetroChat via MSN 8.1 on the Escargo service for user interaction.}
    \end{figure*}

Under the guidance of the design strategies identified, we design and install Retrochat, an GPT-driven chat agent that emulating the way of online expression of 2010 on Chinese social network. The design procedure involving retraving raw conversational data happened on the Tianya Community BBS, developing a chat agent prompted using 2000–2010 dialogue data, crafting its persona to resemble that of a "netizen," hosting a legacy version of MSN software online through a third-party service (Figure ~\ref{fig:Flow v0}).

\setlength{\intextsep}{20pt} 

\subsection{Key Challenges}

Informed by design strategies DS1 and DS2, which specify the time frame for the online conversational data and guide the identification of potential sources for such datasets. A significant challenge that emerged during this process was the retrieval of instant messaging dialogues. Accessing chat histories from instant messaging platforms inevitably raises data privacy concern, as conversations between individual users are generally considered confidential \cite{herbsleb2002introducing}. Consequently, such data is rarely available as open-source datasets, posing significant limitations to their collection and use.

Additionally, from a technical perspective, many platforms that previously hosted conversational data have been discontinued, further complicating access to such datasets. For instance, instant messaging platforms like MSN and ICQ (the predecessor of QQ), as well as social networking platforms such as Kaixin and Renren, are no longer operational \cite{jia2022forgotten}. This challenge extends to the collection of data from BBS forums and open discussions. Many BBS forums, as noted by participants in our formative study phase, have been shut down, making it increasingly difficult to source target data from these platforms.

\subsection{Conversational Data as Corpus}

\subsubsection{Wayback Machine as retrieving tool:}To address the outlined challenges, we utilized the Wayback Machine to collect historical conversational data. Operated by the nonprofit Internet Archive and accessible at archive.org, the Wayback Machine offers access to archived web content dating back to 1996. Its collections have grown significantly over the years, documenting the global rise of websites and online activity from the late 1990s onward \cite{agarwal2022way}. Research leveraging the Wayback Machine has highlighted its value as a large-scale resource for analyzing web information over time. It supports the study of historical trends, the preservation of digital culture, and insights into the evolution of online communication and societal norms, making it an indispensable tool for this type of investigation \cite{arora2016using,ogden2024know}.

\subsubsection{Tianya Club as Data Source:}Armed with this archiving tool and informed by participants' feedback during the formative design phase regarding their use of social networks, we investigated conversational data across three Chinese BBS platforms from 2000 to 2010. Specifically, we analyzed Tianya Club, Sina BBS, and MOP, with Tianya Club emerging as the primary source of conversational data for several reasons:

Significance and Influence: Tianya Club is widely recognized as one of the earliest and most influential Chinese BBS platforms during the first decade of the 21st century \cite{cao2014topics}. Among its spaces, the Zatan and Guanshui boards stood out for their activity, hosting discussions on a wide range of societal topics and fostering broad participation in social events. These boards facilitated in-depth conversations, diverse perspectivesby general internet users and were the incubator of many popular internet slangs, serving as hubs for the creation of popular threads that captured the dynamics of everyday online interactions \cite{chen2014become}.

These attributes make Tianya Club an valuable source for building high-quality corpora, as constructing chat agent personas that authentically reflect the online expressions and characteristics of the time relies on such rich, representative, and interactive content. Moreover, this aligns with the notion that internet expression constitutes a unique form of digital heritage, deeply interconnected with societal opinions, collective mentalities, and cultural values.

\begin{figure*}[h]
        \centering
        \includegraphics[width=0.99\linewidth]{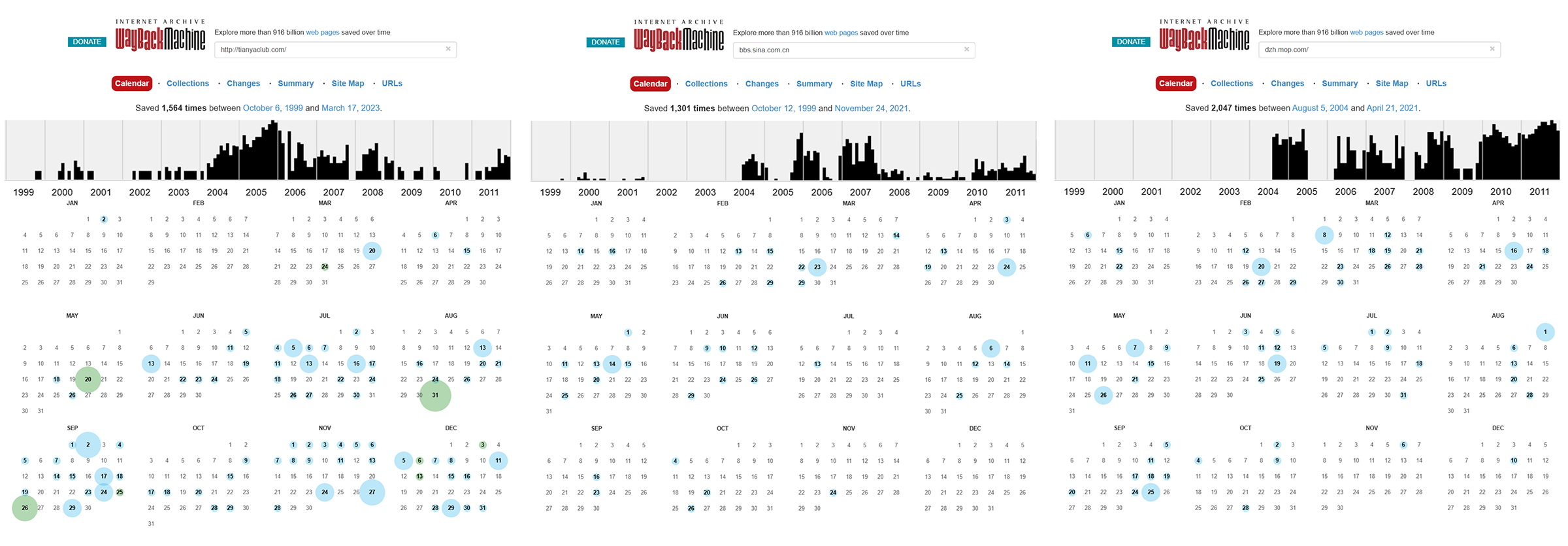}
        \caption{From left to right, the results show the crawler data for Tianya Club, Sina BBS, and MOP, spanning the years 1999 to 2011. Below is an example illustrating the ammount and accessibility in 2005 for each BBS. Blue indicates that the web server's response code for the corresponding capture was successful, while green signifies that the crawler encountered a redirect status.}
        \label{fig:TianyaSinaMop}
        \Description{Data from the Wayback Machine for Tianya Club, Sina BBS, and MOP (1999-2011), highlighting data volume and accessibility in the 2005 archives across three platforms.}
    \end{figure*}

Data Availability and Coverage: We compared the accessibility of archived web pages containing conversational data across the three BBS platforms . Our investigation revealed that MOP had the highest number of archived pages, with 2,047 webpage snapshots, compared to 1,564 for Tianya and 1,301 for Sina BBS. However, MOP's historical pages are limited to the period from 2004 to 2010, with a disproportionate concentration of web crawler data heavily skewed toward post-2007 records. Similarly, Sina BBS exhibits significant gaps, with no data available for the years 2002 and 2003. In contrast, Tianya provides the most extensive coverage, with records preserved from 1999 to 2011. This makes Tianya the richest and most reliable source for our study (Figure ~\ref{fig:TianyaSinaMop}). Eventually, we identified a total of 156,852 characters in the collected dialogue. For all historical data retrieved from Tianya Club, we replaced identifiable information with aliases to ensure anonymity.

\subsubsection{Corpus Building:} To form the main corpus, we generated expression styles using conversational data collected from Tianya. Each style was annotated with examples to help chat agents better understand the appropriate context for employing specific expression styles.

Beyond the main corpus for style learning, we developed a Chinese internet slang index and a collection of nicknames to further enhance the expressive capabilities of our dataset. Using Baidu Wenku, a document search engine, we compiled internet slang terms from the 2000–2010 period. Each slang term was annotated with its year of emergence and formulaic usage patterns, forming the Internet Slang Index Corpus. Additionally, we curated a collection of classic nicknames that were popular among individuals born between 1980 and 1990 and those born between 1990 and 2000. This resulted in the Nickname Collection Corpus, reflecting generational trends in nickname usage.

\subsection{Prompt Engineering}

Following the framework outlined in The Guide to Effective Prompting by Kovari, we implement the Continuous Prompts method combined with a meta-data augmentation approach. This involves enriching the prompts with meta-data, which includes appending relevant details such as instructions, contextual information, and supplementary data to enhance the model's performance on specific tasks \cite{kovari2024chatgpt}. Such a strategy is particularly beneficial for improving the model's understanding and enabling it to generate more accurate and contextually appropriate responses, especially when the contextual information provided is detailed and rich.

In this study, the metadata comprises two key components: (1) a main corpus of processed chronogical conversational data from Tianya Club, integrated into GPT's knowledge base to enhance its understanding of Chinese internet expressions and communication styles from the 2000–2010 period; and (2) a Chinese Internet Slang Index, documenting popular slang and specific formulaic expressions for each year. (3) a nick name collection from 2000-2010. The configuration of the GPT-4 prompts mainly consists role description, conversational style, and response formate.

\setlength{\intextsep}{20pt} 

\subsubsection{Persona Description:}This part define the role of the chat agent to be an 2010 Chinese social network netizen. the agent aims to represented a persona typically informed by conversational patterns and cultural norms documented in the corpus we building Maintain consistent identity and authenticity throughout the conversation to create a historically immersive and realistic interaction.
    \begin{quotation}
    \textit{"You are a netizen of Chinese social networks representing 2000-2010 online expression. Your character should authentically embody the typical traits and behaviors of netizens from that period, as informed by the /main corpus/, which documents conversational patterns among users of that time. Adopt a nick name inspired by the /nick name/, ensuring it reflects the cultural context of 2010. Consistently maintain this identity during the conversation. Respond naturally as a human would from that era, avoiding overly detailed or chatbot-like solutions. It is acceptable to express uncertainty or admit a lack of knowledge when appropriate, also you can starting a topic appropriately."}
    \end{quotation}

\subsubsection{Conversational Style:}This section outlines the stylistic approach using metadata from conversational data and a collection of slang. According to our design strategy, DS4, the use of slang should peak around 2010 and gradually decline toward 2000, mirroring historical trends. Expressions must align with the linguistic and cultural context of the era, avoiding modern elements, to authentically capture the conversational style and cultural nuances of the 2010 online environment.
    \begin{quotation}
    \textit{"Simulate the online communication style of Chinese netizens representing the 2000 to 2010 era, referencing the /main corpus/ for instruction on tone, emotional expression, and conversational flow.  Integrate expressions and terms from the /slang index/ in a contextually appropriate and historically accurate manner. Balance slang usage to ensure it enhances the natural flow of conversation, aligning with its original meaning and cultural context. The frequency of slang should peak around 2010 and progressively decline from 2009 to 2000, reflecting historical trends. Ensure that all expressions adhere to the linguistic and cultural norms of the period, avoiding any inclusion of modern terminology or stylistic elements that could disrupt the immersion. The goal is to authentically replicate the online conversational style and cultural nuances of the 2000–2010 era."}
    \end{quotation}

\subsubsection{Response Format:}This section reinforces the chat agent's character as a netizen and underscores human-like conversation. It also provides a record of past dialogues, enabling the agent to maintain continuity and deliver coherent, contextually relevant interactions with the user \cite{calderwood_spinning_2022}. 
    \begin{quotation}
    \textit{"Vary your chat styles as a netizen, using a human-like way of conversation by sometimes asking questions and other times sharing information. Follow the typical string-based response format, introducing stylistic variations to maintain the naturalness and authenticity of interactions. Use prior discussions from /conversation/ as context to ensure continuity and coherence. Keep responses dynamic and engaging, adapting to the evolving context of the conversation."}    
    \end{quotation}


\subsection{Artefact Installing}

\setlength{\intextsep}{20pt} 
    \begin{figure*}[h]
        \centering
        \includegraphics[width=0.99\linewidth]{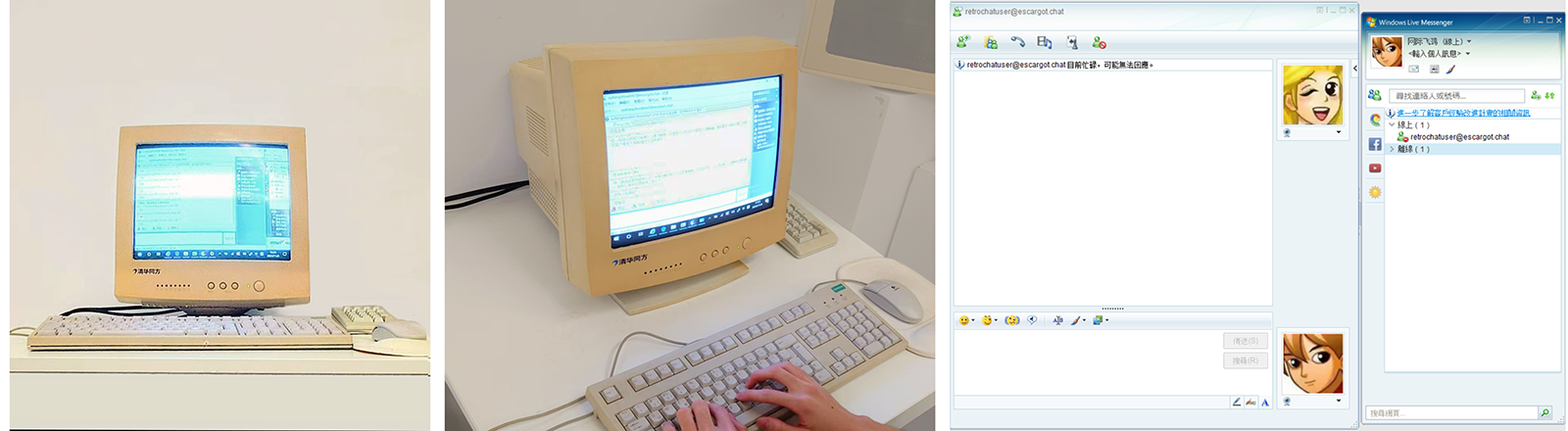}
        \caption{The left figure shows the overall setup of the environment, the middle figure illustrates how users engage and interact with the artifact, and the right figure displays the legacy MSN 8.1 interface where participants chat with the agent.}
        \label{fig:SetUp}
        \Description{Physical setup of RetroChat, including the environment, user interaction, and MSN 8.1 interface used for chatting.}
    \end{figure*}

Our design implementation aimed to authentically replicate an immersive retro chatting experience involves setting up environment of hardware and deploying a legacy version of MSN Messenger. To achieve this, we used Escargot, which, to the best of our knowledge, is the accessable service for restoring outdated SNS software, supporting all legacy versions of MSN and WLM clients. Based on our The benefits of this approach were significant, as it not only revived the retro user interface but also restored the interactive functionality of Chatting system of the era, enabling authentic user engagement. Due to the lack of exact data, we ran the 2007 version of MSN 8.1, identified as the most popular version of MSN in China during that time based on time-based speculation \cite{asia_top_2015}. Regarding presenting of chat agent, we select the iconic avatar from early Chinese social network as DS4 suggested. The hardware setup included a Qinghua Tongfang CRT monitor, a Nixdorf 120 mechanical keyboard, a Cherry G80-3700 Keypad, and a Logitech M-S34 ball mouse, depicting a fully period-accurate emulation of the retro online chat environment (Figure ~\ref{fig:SetUp}).

%% file: sections/05-Method.tex
\subsection{Recruitment Method}
We recruited 18 participants through online platforms, including using WeChat Moments and Redbook. Our study aimed to explore how GPT-driven interactive designs can preserve the experience of chatting on retro Chinese social networks. Since typing and communicating in Chinese were essential components of the experiment, we specifically required participants who preferred using the Chinese language. Participants were compensated with 50 Chinese Renminbi or the equivalent in Hong Kong dollars upon completing the experiment and the subsequent interview.

\begin{itemize} \item Consent: Participants provided informed consent prior to their involvement in the study. The consent form clearly detailed the research objectives, procedures, potential risks, and benefits. It also highlighted the voluntary nature of participation and emphasized the participants' right to withdraw at any point without any consequences. \item Anonymity: Participant privacy and anonymity were strictly upheld. No personally identifiable information was collected, and all necessary measures were taken to prevent the acquisition or retention of any sensitive data that could reveal participants' identities. \item Ethics Approval: The study received approval from the University's Institutional Review Board (IRB), ensuring compliance with ethical standards for human subject research. \item Ethical Conduct: The research adhered to core ethical principles, including respecting participants' rights, minimizing potential risks, and maintaining strict confidentiality and security of all collected data. \end{itemize}

\subsection{Participants}
Our study included 18 participants, comprising 8 females and 10 males, aged between 21–30 and 40–50 years. Among the participants, 8 began using online social networks between 1995 and 2000, 9 started between 2001 and 2005, and 1 began in 2006 or later. Regarding familiarity with 2000–2010 internet slang, 9 participants considered themselves as familiar and actively practicing these expressions, 7 reported familiarity without active usage, and 2 indicated no familiarity. Detailed as below.

\begin{table*}[t]
\centering

\resizebox{0.98\textwidth}{!}{%
\begin{tabular}{|l|l|l|l|l|l|}

\hline
\textbf{ID}  & \textbf{Age} & \textbf{Gender} & \textbf{Education Level}    & \textbf{First time use SNS?} & \textbf{Past Online Expression Familiarity?} \\ \hline
\textbf{p1}  & 31-40        & Male            & Junior high school or below & 1995-2000                                                  & Practice                                            \\ \hline
\textbf{p2}  & 21-30        & Male            & Master's degree             & 2001-2005                                                  & Knowledgeable                                       \\ \hline
\textbf{p3}  & 40 and above & Male            & Bachelor's degree           & 1995-2000                                                  & Practice                                            \\ \hline
\textbf{p4}  & 31-40        & Female          & Bachelor's degree           & 1995-2000                                                  & Practice                                            \\ \hline
\textbf{p5}  & 21-30        & Male            & Master's degree             & 2001-2005                                                  & Practice                                            \\ \hline
\textbf{p6}  & 21-30        & Female          & Master's degree             & 2001-2005                                                  & Knowledgeable                                       \\ \hline
\textbf{p7}  & 21-30        & Female          & Doctoral degree or above    & 2001-2005                                                  & Practice                                            \\ \hline
\textbf{p8}  & 21-30        & Female          & Bachelor's degree           & 2001-2005                                                  & Knowledgeable                                       \\ \hline
\textbf{p9}  & 40 and above & Female          & Bachelor's degree           & 1995-2000                                                  & Practice                                            \\ \hline
\textbf{p10} & 21-30        & Female          & Master's degree             & 2001-2005                                                  & Knowledgeable                                       \\ \hline
\textbf{p11} & 21-30        & Male            & Master's degree             & 2001-2005                                                  & Knowledgeable                                       \\ \hline
\textbf{p12} & 40 and above & Male            & Bachelor's degree           & 1995-2000                                                  & Practice                                            \\ \hline
\textbf{p13} & 40 and above & Male            & Junior high school or below & 1995-2000                                                  & Knowledgeable                                       \\ \hline
\textbf{p14} & 21-30        & Female          & Master's degree             & 2001-2005                                                  & Unknowledgeable                                     \\ \hline
\textbf{p15} & 21-30        & Male            & Master's degree             & 2001-2005                                                  & Unknowledgeable                                     \\ \hline
\textbf{p16} & 21-30        & Female          & Bachelor's degree           & 2006 and above                                             & Knowledgeable                                       \\ \hline
\textbf{p17} & 31-40        & Male            & Bachelor's degree           & 1995-2000                                                  & Practice                                            \\ \hline
\textbf{p18} & 31-40        & Male            & Master's degree             & 1995-2000                                                  & Practice                                            \\ \hline
\end{tabular}%
}
\caption{Summary of participants in the experiment, including demographic information.}
\Description{Summary of 18 participants in the experiment.}
\end{table*}

\subsection{Experimental Procedure}
After obtaining confirmed consent from each participant, the experiment was initiated. The experiment procedure consisted of three parts: (1) the pre-task phase, where participants' information was gathered, (2) the main task, which involved chatting with the Retro Chat system, and (3) the post-task interview, which included a retrospective think-aloud session and open-ended questions (Figure ~\ref{fig:Experimental Procedure}).

During the pre-task phase, we first collected participants' demographic information, including age, gender, educational level, and the year they first started using SNS. Next, participants completed a familiarity assessment related to 2000–2010 Chinese SNS and past Chinese internet language. In this stage, they were shown snapshots of BBS and legacy SNS platforms to help contextualize what we meant by "2000–2010 Chinese SNS." They were then asked to complete an 8-item questionnaire using a 5-point Likert scale to assess their experience with early social networks. Following this, participants self-reported their past knowledge and practice of Chinese online expressions using the following categories:
\begin{itemize}
    \item \textbf{Level Unknowledgeable}: Not familiar with 2000–2010 way of expression.

    \item \textbf{Level Knowledgeable}: Familiar with 2000–2010 way of expression but do not actively use it.

    \item \textbf{Level Practice}: Familiar with 2000–2010 way of expression and actively use it.
\end{itemize}
Finally, participants were asked an open-ended question: How did you perceive online chatting during that time compared to how you perceive it today?

\setlength{\intextsep}{20pt} 
    \begin{figure*}[h]
        \centering
        \includegraphics[width=0.99\linewidth]{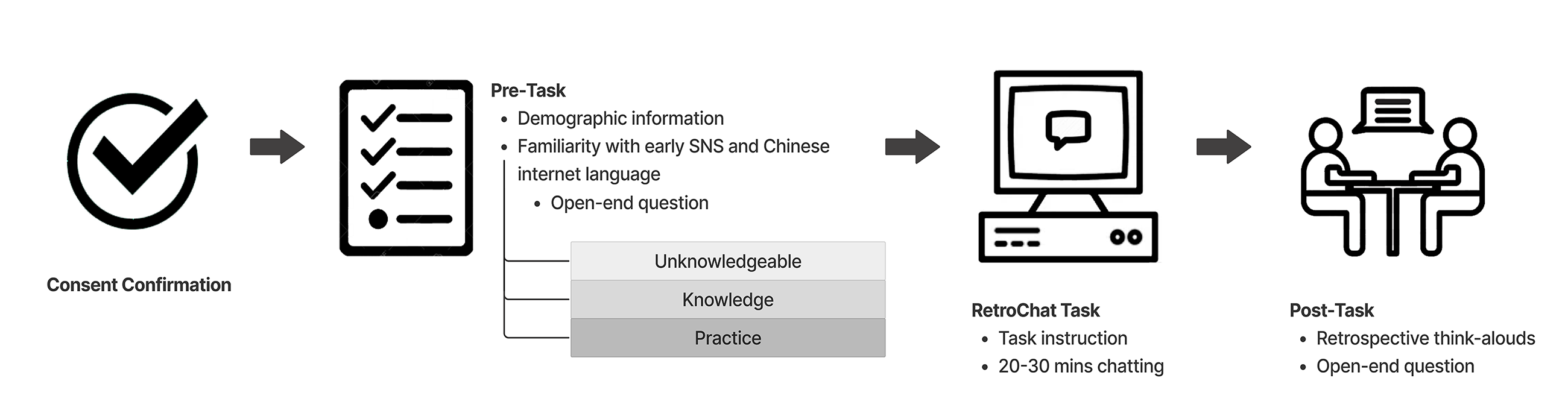}
        \caption{Experimental Procedure: After obtaining consent, the study is conducted in three stages. The first stage is the Pre-task, which involves collecting participant information, including their familiarity with Chinese SNS from 2000-2010. The second stage is the Chatting Task, where participants engage with RetroChat. The final stage is the Post-task, a retrospective interview featuring open-ended questions.}
        \label{fig:ExperimentalProcedure}
        \Description{Experimental procedure: Pre-task (participant details collection), Chatting task (interaction with RetroChat), Post-task (retrospective interview)}
    \end{figure*}

After the pre-task phase, participants were invited to sit in front of the setup and given time to familiarize themselves with it to assist those unfamiliar with using a ball mouse or a vintage keyboard layout. Once they were comfortable, we launched RetroChat and ask participant to engage in chatting as if they meet a netizen:
    
    \begin{quotation}
    \textit{"You are in the early days of Chinese social networks, surfing the internet. You encounter an online netizen and start a conversation. There are no restrictions on the topics you can discuss—whether it’s daily life, trending topics, or anything else. Simply engage in a conversation and chat naturally about anything that comes to you."}
    \end{quotation}

To ensure a natural conversational environment and minimize external pressure that could influence the chat flow, participants were not informed of the time limit in advance. Existing research underscores the importance of studying conversational behaviors in naturalistic settings to capture authentic interactions \cite{clark1991grounding, bavelas2000listeners, patton2014qualitative}. Although no time constraints were communicated beforehand, the maximum duration for each participant was set at 30 minutes. If a participant exceeded this limit, they were gently informed that they could conclude the conversation at their own pace.

Once participants completed the chatting task, the study proceeded to the interview phase, where we employed the Retrospective Think-Aloud (RTA) technique. The rationale for using RTA was to preserve the natural flow of conversation, as participants were already engaged in communication, primarily language processing, during the RetroChat task \cite{noushad2024twelve}. The procedure of the RTA is as follows: After completing the task, participants were presented with their chat history with the chat agent. They were then asked to verbalize their thoughts, emotions, and feelings experienced during the interaction. 

\subsection{Qualitative Analysis}

We collected chat history from 18 participants interacting with RetroChat, totaling 50,476 characters. Additionally, we recorded and transcribed 232 minutes of interview data. Participants were informed that all chat history during the task and interview recordings would be securely stored on a local drive, accessible only to researchers. In compliance with IRB regulations, the data was permanently deleted upon completion of the analysis.

For data analysis, two authors independently analyzed the chat history and interview transcripts. We began with an initial review of all chat logs to extract and document conversational behavior patterns as textual descriptions. These descriptions served as a foundation for guiding the analysis of the interview transcripts. For the interview transcripts, we conducted thematic analysis \cite{boyatzis1998transforming, clarke2017thematic} to identify recurring codes and themes. The initial codes and themes generated by each researcher were then aggregated and refined through multiple rounds of iterative discussions among the research team. Throughout this process, participant narratives and quotes were incorporated into the main themes to provide supporting evidence and enhance the depth of our analysis.

%% file: sections/06-Result.tex
\subsection{Adapting Chatting Styles to Past Internet Expressions}

One notable phenomenon observed is that participants spontaneously adjusted their language to align with the RetroChat agent’s style. Specifically, among the 18 participants, we found that 7 incorporated past internet slang, neologisms while chatting various topics with the chat agent.

For example, P3 used the internet slang \begin{CJK*}{UTF8}{gbsn}“天上飘来五个字，这都不是事儿”\end{CJK*}, which literally means "A phrase floated down from the sky: This is no big deal," to express a relaxed, easygoing attitude when discussing working conditions with the chat agent. P7 was observed using a combination of slang terms and numeric abbreviations during the conversation. When chatting about celebrities, P7 used "818" to represent \begin{CJK*}{UTF8}{gbsn}"扒一扒"\end{CJK*}, meaning "gossip" in English. When the topic shifted to how people used image editing software to make themselves look more attractive, P7 humorously and mockingly quoted a once-popular internet phrase:\begin{CJK*}{UTF8}{gbsn}"捂一只眼睛自拍,45度角的忧伤你不懂"\end{CJK*}, which translates to "Cover one eye, tilt your head at a 45-degree angle—the sorrowful, cool aesthetic, you just don’t get it." 

We observed participants using neologisms during their conversations. For example, when discussing the experience of searching for a favorite movie, P12 used the slang term \begin{CJK*}{UTF8}{gbsn}"杯具了"\end{CJK*} (literally "cup" but used as a pun for "tragedy") to express frustration with the search process, commenting on how some online search engines were not as convenient as expected. Similarly, P7 also used \begin{CJK*}{UTF8}{gbsn}"杯具"\end{CJK*} when talking about how some SNS platforms have been shut down, resulting in the loss of online photo archives. When discussing concert experiences with the chat agent, P10 learned that the agent had not attended any recent live events. In response, she used the phrase \begin{CJK*}{UTF8}{gbsn}"蓝瘦"\end{CJK*}, a transliteration of \begin{CJK*}{UTF8}{gbsn}"难受"\end{CJK*} (meaning "uncomfortable" or "upset"), contextually expressing a sigh of disappointment in reaction to the agent’s response. P17 used the phrase \begin{CJK*}{UTF8}{gbsn}"打酱油"\end{CJK*} twice—once to humorously describe himself as an unimportant or peripheral participant when discussing his role in team-based work and again to characterize his gaming skills when talking about gameplay.

As the chat agent incorporated past internet expressions, such as referring to user gender with "GG" and "MM" (popular internet slang in the 2000s representing "boy" and "girl," respectively), participants naturally adapted to this linguistic style. When the chat agent fictionalized itself as a "GG from Northeastern China," P13 naturally continued referring to it as "GG" later in the conversation when discussing Northeastern cuisine. A similar pattern occurred with P18 — when the chat agent used "GG" to refer to the user, P18 also began using the same substituted term to address the agent during their conversation. This suggests that participants not only picked up on the agent’s linguistic cues but also mirrored its style of addressing them. 

\begin{table*}[]
\begin{tabular}{|l|l|l|l|l|l|}
\hline
\textbf{ID}  & \textbf{Age} & \textbf{Gender} & \textbf{First time use SNS?} & \textbf{Familarity} & \textbf{Used expressions}  \\ \hline
\textbf{p10} & 21-30        & Female          & 2001-2005                    & Knowledgeable                                       & Popular Words              \\ \hline
\textbf{p12} & 40 and above & Male            & 1995-2000                    & Practice                                            & Slang/Popular Words        \\ \hline
\textbf{p13} & 40 and above & Male            & 1995-2000                    & Knowledgeable                                       & Abbreviation               \\ \hline
\textbf{p17} & 31-40        & Male            & 1995-2000                    & Practice                                            & Slang                      \\ \hline
\textbf{p18} & 31-40        & Male            & 1995-2000                    & Practice                                            & Abbreviation/Popular Words \\ \hline
\textbf{p3}  & 40 and above & Male            & 1995-2000                    & Practice                                            & Slang                      \\ \hline
\textbf{p7}  & 21-30        & Female          & 2001-2005                    & Practice                                            & Slang/Abbreviation         \\ \hline
\end{tabular}
\caption{The table indicates the seven participants who incorporated past online expressions, categorizing the types of expressions used. It also includes their first-time use of SNS and their familiarity with Chinese social networks from 2000 to 2010.}
\Description{List of 7 participants using past online expressions, categorized by expression type, SNS usage, and familiarity with Chinese social networks from 2000-2010.}
\end{table*}

We found that among these seven participants, five identified themselves as well-versed in internet expressions and online culture from 2000 to 2010 and had actively used them during that time. The remaining two participants reported having knowledge of past internet culture and online expressions but did not actively engage with them. Regarding social networking service (SNS) usage history, six participants traced their first SNS experience back to 1995–2000, while one participant reported first using SNS between 2001–2005. This linguistic adaptation suggests that the design resonates with participants who are familiar with and had early exposure to internet culture, as reflected in their ability to re-engage with past online linguistic styles.

\subsection{RetroChat as a Memory Kaleidoscope}

During participants' conversations with the RetroChat Agent, we observed that they naturally gravitated toward various topics related to their past experiences. While these topics were not always the main focus of the conversation, they frequently emerged across different participants. This phenomenon is particularly interesting because, in designing the prompting system, we deliberately avoided steering the conversation by maintaining the persona of a netizen from that era while promoting free and organic discussions. 

\subsubsection{Recollections of Early Online Life}

The chat history between participants and the RetroChat Agent reveals recurring discussion patterns centered around early internet experiences and how people sought entertainment in the digital space. Certain frequently mentioned online activities appeared across different conversations, suggesting a collective memory of early Chinese online culture that reflects shared experiences from that era.

Participants recalled various aspects of their early online activities. For example, P6 shared her experience of searching for emoji stickers online:\textit{"One thing I used to do all the time was go online and download pretty emoji stickers by myself. It’s different from now, where most people download emoji packs for humor. Back then, my friends and I would search for delicate and beautifully designed emoji stickers. It was like a little girl's dream of growing up into a sophisticated city woman, hahaha!"}

P7 reflected on how visiting and interacting with friends’ personal pages was an essential part of online social culture, contrasting it with modern social media norms: \textit{"Yeah! I used to hate when people visited my space without leaving a footprint. Xiaohongshu, Weibo, and WeChat Moments have taken over everything and formatted all online interactions. Back then, leaving a footprint in someone's space was basic courtesy! Yeah, but my old photo albums were way too shamate."} 

P18 described the excitement of using P2P resource-sharing sites, emphasizing how information was harder to find back then, making the discovery process more thrilling:\textit{"There was a time when I was really into P2P resource-sharing sites. I mean, now information is everywhere, but I remember back in the day, finding those online resources—like open courses—felt like digging for gold."}

Among those online entertainment bring in the conversation, a particular category of web-based social games was repeatedly mentioned by different participants. These games, primarily farm-themed social simulations, allowed players to manage virtual farms, cultivating, irrigating, and harvesting crops while interacting with friends' farms \cite{kow2012designing}. Players could cooperate by weeding and watering each other’s crops or compete by stealing matured crops, directly affecting in-game income. A personal message board tracked SNS friends’ visits and recorded their interactions.

P7 shared a vivid memory of playing Happy Farm during their chat with the agent, recalling both the mechanics and a personal anecdote:\textit{"Everyone had their own little virtual farm where they could grow things like corn or Chinese cabbage, and we’d set alarms to 'harvest' them on time. If you were just a few minutes late, those 'crop-stealing maniacs' would have already paid you a visit—leaving nothing behind, not even a single leaf! It drove me crazy! I was thinking about harvesting my crops even during class and almost got caught red-handed by my homeroom teacher!"} Similarly, 

P16 explored whether the chat agent recognized the concept of "stealing crops" and attempted to share gaming techniques. During the conversation, P16 asked the agent questions such as: \textit{"Do you know the technique people used to sneak onto their friends' farms? You have to choose the right time to steal vegetables and leave a time gap."} In the later interview, P16 elaborated on their memory flashback during the chat, expressing deep nostalgia for these social games and describing them as an immersive, alternative digital world: \textit{"you know, logging in every day to farm — it was like having a second life. There are still games like this now, but I just can't seem to find people to play with anymore, at least not in the same way as back then."}

P2 and P17 also briefly mentioned their experiences with online gaming, though without extensive descriptions. P2 recalled how gaming introduced them to a broader social network:\textit{"...I started by watching her (P2's older cousins) play and eventually joined in. Turns out, I really enjoyed it—games like Mole's World and Fly for Fun—and even made some friends along the way..."} The spontaneous emergence of these gaming-related discussions in free conversation suggests that these ludic experiences have left a lasting imprint, forming a collective memory for those who lived through that era of digital life.

\subsubsection{Memory Flashbacks Extending Beyond Online Spaces}
We also found that some participants spontaneously recalled real-life experiences from their youth, often sharing seemingly trivial yet somehow memorable personal anecdotes. Through interaction, these memories triggered broader autobiographical recollections from earlier stages of life. For instance, P2 reminisced about childhood memories of playing with fireworks when discussing plans for the New Year. P6 and P16 remained engaged in nostalgic reflection for an extended period, sharing vivid and detailed personal stories.

\setlength{\intextsep}{20pt} 
    \begin{figure*}[h]
        \centering
        \includegraphics[width=1\linewidth]{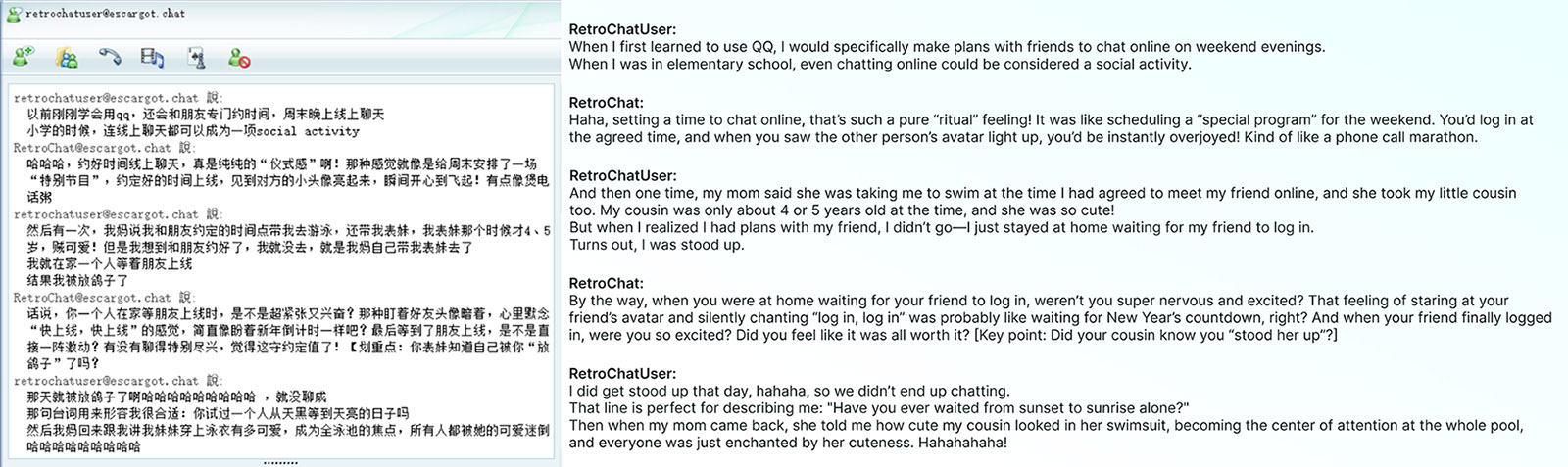}
        \caption{Detailed chat history of P6 recalling a childhood memory of being stood up by her mother and sister, sharing this experience with the chat agent during the conversation.}
        \label{fig:P6}
        \Description{Chat history of Participant 16 recalling a local street food and a humorous incident from her school years.}
    \end{figure*}

P6 recounted a memorable incident: one weekend, despite her mother's plan to take her swimming with her adorable 4- or 5-year-old cousin, she chose to stay home for her scheduled online chat. Unfortunately, her friend never logged in, leaving her waiting alone. RetroChat playfully compared this anxious wait to the excitement of a New Year’s countdown and even teased whether her cousin had noticed she was stood up (Figure ~\ref{fig:P6}).

\setlength{\intextsep}{20pt} 
    \begin{figure*}[h]
        \centering
        \includegraphics[width=1\linewidth]{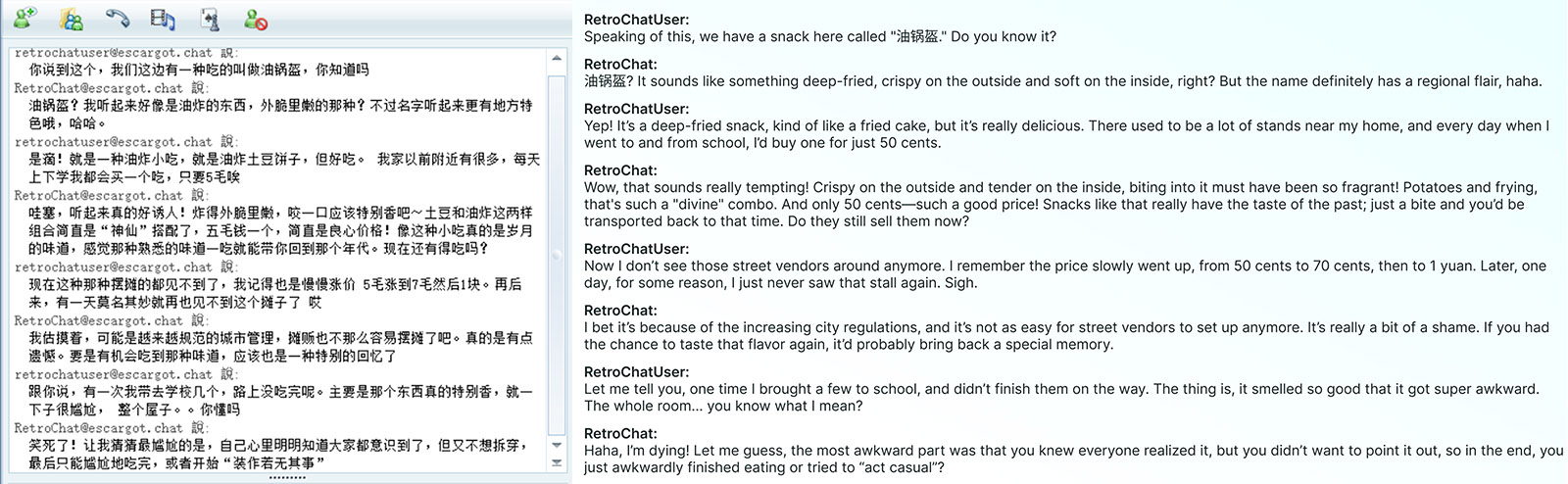}
        \caption{Detailed chat history of P16 sharing a nostalgic story about a local street food she enjoyed during her school years and a humorous incident related to it.}
        \label{fig:P16}
        \Description{Detailed chat history of P16 sharing a nostalgic story about a local street food she enjoyed during her school years and a humorous incident related to it.}
    \end{figure*}

Similarly, P16 vividly recalled a local snack. She explained that as a child, she would buy this deep-fried delicacy from street vendors for just 50 cents every day on her way to and from school. Over time, rising prices and stricter city regulations led to the disappearance of these vendors, leaving only nostalgic memories. P16 also shared a humorous incident where the snack’s irresistible aroma created an awkward moment at school (Figure ~\ref{fig:P16}).


\subsection{RetroChat as a \begin{CJK*}{UTF8}{gbsn}"树洞"\end{CJK*} (ShuDong) for Expression and Confiding}

We observed that several participants perceived RetroChat as a safe space for sharing personal thoughts and emotions without fear of judgment. For example, P5 expressed concerns about body image, stating, \textit{“I want to lose weight to reach 160–165 pounds, and it would be ideal to develop defined abdominal muscles, as I believe physical fitness is genuinely appreciated.”} Additionally, some participants voiced dissatisfaction with their work-life balance. When discussing workplace challenges, P3 remarked, \textit{“I am working on a big project now—it feels like all I do is perform menial tasks every day, with no change or progress.”} Similarly, P2, reflecting on their role as a teaching assistant, stated, \textit{“I have been working overtime for a week straight, and my boss still asks for more—I am exhausted.”} P15 also raised concerns about the rapid pace of technological change, questioning the practical and commercial value of their work: \textit{“I feel that what I’m doing is just for fun and lacks real impact or commercial value. If I really want to make something meaningful, I need to develop an app that truly makes a difference.”}

This observation suggests a potential application for our design—functioning as a Shudong \begin{CJK*}{UTF8}{gbsn}(树洞)\end{CJK*}, a metaphorical "tree hole" where users can freely express emotions, share personal thoughts, and engage in deeper reflections that might otherwise remain unspoken \cite{zhang2020placing}. Our interviews exploring participants' perceptions of the chat agent's persona may offer some explanations. P2 described the agent as sincere, saying, \textit{"It feels like talking to an old friend I can trust,"} and added, \textit{"This kind of communication has become uncommon on the Internet."} P5 bring a reflection on how the concept of "netizens" has evolved over time, stating, \textit{"This really reminds me of how different the idea of 'netizens' used to be. Back then, people embraced the term—it was so easy to chat with a stranger and even add them as a friend. Now, it feels like everyone is just 'online,' and the way we express ourselves has become much more polished."}

Although P6 did not use RetroChat as a platform for complaints or confessions, she described the agent’s personality as buoyant, frank, and easygoing, though at times a bit too youthful—\textit{"like a stereotypical high school student."} Similarly, P8 remarked that the chat agent’s persona reminded her of a classmate from her past. Some participants expressed opposing views when discussing the chat agent’s persona. P13 felt that the agent’s expressions were a bit exaggerated, commenting, \textit{"I get the sense of what it's going for, but it’s a bit over the top. Some of these terms were only used naturally in specific contexts."} P9 also found the interaction somewhat unnatural, stating, \textit{"It feels contrived and warm—the chatbot responds too quickly, which disrupts the natural flow of conversation."}

\subsection{Y2K Aesthetic Revisiting}

When asked about the emotions related to their interaction experience, eight participants invoked the word "nostalgia" to describe their engagement, using language that highlighted its aesthetic allure. In particular, P7 and P16 recounted vivid flashbacks of their youth, which surfaced like cherished, imagistic, cinematic memories throughout the chat. Notably, P6 and P7 described their feelings as "warm" and "arousing," while P8 experienced a sense of melancholy when discussing the emotions evoked by the artifact. Interestingly, both P6 and P8 used the term "dissociation" to characterize their experience. P6 was able to elaborate in more detail: \textit{"...so I'm talking to someone from a past era, but at the same time, I can bring up today's topics, and they respond in their own retro way. That's when I realized how different things used to be—even something as simple as chatting had a different feel back then. It's a unique aesthetic of Y2K, but also kind of surreal."} P7 connected the experience to a broader artistic movement, stating: \textit{"The way it interacts with me reminds me of Chinese "Nostalgiacore" or "Oldcore" internet artwork—like, the whole aesthetic with old PCs and that  2000ish setup."} These reflections suggest that the artifact's visual and interactive design inspired deeper thoughts about digital culture and memory, evoking complex emotions and thus highlighting its potential artistic value.

%% file: sections/07-Discussion.tex

Our study of RetroChat provides valuable insights into the role of AI conversational agents in preserving past digital experiences and facilitating chatting interactions. The findings highlight how users engaged with the chat agent in a way that not only mirrored past online linguistic styles but also triggered broader autobiographical recollections, enabling in-depth reflection after the interaction. In this discussion, we reflect on the implications of these findings for digital heritage preservation, social computing, as well as on the limitations and future directions of this research.

Our findings demonstrate that RetroChat effectively elicited in-depth reflections and memories associated with early Chinese internet culture. Participants, particularly those familiar with retro Chinese online culture, spontaneously adopted past internet expressions, including slang and stylistic elements characteristic of early online interactions. This suggests that AI-powered conversational agents designed to emulate historical digital vernacular can serve as an experiential approach to digital heritage preservation, resonating most with individuals who have lived through these digital eras. This aligns with the growing interest in digital cultural memory, where digital heritage is increasingly viewed as an interactive process that enables individuals to engage with memories by sharing their own experiences and exploring those of others \cite{adiba2025exploring, namiceva2018ai, pataranutaporn2023living}. 

Furthermore, participants’ engagement with the chatbot extended beyond online experiences, prompting recollections of real-life events from their youth. Such a phenemenon suggests that experiential interactions, particularly those embedded within specific sociocultural contexts, can serve as powerful memory cues that transcend the digital realm, illustrating the deep connection between online engagement and personal history within its era.



\subsection{Experiential Approach toward Broader Preservation Action}
While our study focuses on a specific user group within the context of Chinese online culture—particularly the fading internet expressions from the 2000s to 2010s—we argue that this experiential approach has broader applicability. Researchers examining other user groups or cultural contexts may also find such methods beneficial. 

For instance, MySpace was once a dominant social networking platform in the mid-2000s, hosting millions of users from diverse backgrounds and accumulating vast amounts of textual content and user-generated data \cite{torkjazi2009hot}. However, due to a faulty server migration, MySpace lost all content uploaded prior to 2016 \cite{haliburton2021quick}. This failure resulted in the loss of a significant portion of early internet content, along with valuable insights into how early netizens crafted their online identities and the kinds of information they chose to share. By combining internet archiving techniques with AI-driven dialogue systems, researchers can begin to reanimate such lost digital cultures. These systems offer the possibility of experiential interaction—allowing users to sense what it might have felt like to engage with a platform like MySpace during its prime.

Moreover, the AI-driven approach presented in this study can be meaningfully extended to broader cultural preservation efforts, particularly in contexts where historical data and human-like communication are essential. Researchers and practitioners in cultural heritage preservation have increasingly adopted interactive methods—such as virtual reality simulations or gamified experiences—to revisit and re-experience historical scenarios \cite{hutson2024combining, nortoft2024gamifying, kindenberg2025role}. These efforts often require conversational agents capable of authentically representing individuals within specific temporal and sociocultural contexts \cite{pataranutaporn2023living, colucci2024automated}. Our study demonstrates how prompt engineering can enable AI agents to embody historically grounded digital personas that resonate with users from the corresponding era. This technique may be especially valuable for scholars seeking to create engaging, interactive characters for use in design, education, or public engagement.

\subsection{Conversational AI for Studying Real-Time Digital Behavior on Historical Platforms}

Another key contribution of our work is that it offers insight into an alternative methodological approach that moves beyond textual analysis and traditional user studies to investigate historical social networking. Our study suggests that interactive chat-based systems can serve as valuable tools for examining user behavior. By observing how participants engaged with RetroChat, we identified patterns indicating that past social and ludic experiences—as well as online conversations—leave an enduring imprint on users, forming collective digital memories. This methodology is related to recent work using natural conversation in games to infer player offline behaviors \cite{zhang_can_2025}.

Traditional approaches to studying past online behavior often rely on static archived records in text and media or self-reported recollections, both of which may fall short in capturing the dynamic, conversational nature of these interactions and their contextual nuances \cite{ogden2022everything, lomborg2012researching}. RetroChat, however, offers a simulated, experiential approach that enables users to interact  with past digital cultures, and actualizing researcher to observing user's real time interaction, thus allowing for a richer and more direction exploration of those interactions. 

This approach facilitates the investigation of research questions such as how contemporary users engage with past online personas and historical digital environments, as well as how they perceive, interpret, and reflect on these experiences. Future research could extend this chatbot-driven methodology to a range of historical digital contexts—such as retro gaming communities, early blogging cultures, or cyber subcultures—to examine the evolution of online social behaviors, shifts in public discourse, and the ways in which digital spaces shape identity formation.

Researchers may also leverage the real-time nature of this approach in experimental designs. While the present study employed retrospective think-aloud protocols and post-interaction interviews—which revealed users’ reflective engagement—we contend that this method also holds potential for capturing real-time, in-task data. As users actively participate in interactive conversations, researchers can incorporate additional modalities—such as biometric measures—to enable a multimodal analysis of user behavior and emotional engagement \cite{shibuya2022mapping}. 




\subsection{Limitation and Future Plan}

The sample size in the formative study phase was limited, and participant recruitment primarily relied on the researchers’ social networks, which may have introduced bias. Future investigations could benefit from expanding both the diversity and size of the sample to enhance the generalizability of the findings and to inform a more comprehensive design strategy. Among participants who performed the task, the sample appeared to be limited to individuals with prior experience using early Chinese social networks, which may not fully represent broader internet user demographics. This sampling outcome may suggest that a significant portion of Chinese netizens still retain awareness of early online expressions, given that China first gained internet access in 1994. Since our primary objective was to develop an authentic and representative interactive form of digital preservation, this participant pool was appropriate for our initial exploration. However, an important direction for future research would be to investigate how younger generations—who lack direct experience or familiarity with early internet culture—engage with such systems.

Additionally, we introduced a physical setup using an old computer system, following Design Suggestion 6 (DS6) identified during the formative study, to better mimic the time-specific context of early internet use. This design strategy was adopted based on our consideration of physical context as an integral part of the user experience. However, this raises questions about the generalizability of such a system. To address this limitation, future investigations could test the system on contemporary computer setups to further explore its adaptability and relevance across different technological environments.

Beyond that, while participants generally reported positive engagement, some noted during interviews that RetroChat’s interaction flow could be improved for a more natural and context-sensitive conversational experience. Currently, the system does not fully adapt dynamically to user inputs beyond certain linguistic features, particularly in scenarios where the historical context of online expressions plays a crucial role. Future work could focus on iterative testing to refine conversational quality. Additionally, as LLM technology advances, we plan to explore models with greater specialization in the Chinese language \cite{bi2024deepseek,li2024impact}, improving contextual awareness and linguistic accuracy.

Finally, while this study primarily examined conversational interactions, the design of the contextual physical environment was primarily focused on replicating the PC experience of the time, following a grounded design strategy. However, our findings suggest that digital preservation may also hold artistic value beyond historical accuracy. Given recent work in participatory GenAI art practice \cite{lc_together_2023, lc_human_2023, lc_time_2024, zhou_eternagram_2024-1}, an alternative direction of the present work could explore additional design elements, such as visual aesthetics and interface affordances, to further enhance immersion in past digital experiences and expand the scope of engagement.

%% file: sections/08-Conclusion.tex
As social networks rapidly evolve, shifting linguistic expressions and digital norms make it increasingly difficult to preserve past digital experiences. Traditional web archiving captures only static content, failing to account for the dynamic nature of digital conversations. Our study demonstrates the design of AI-driven conversational agents for preservation through re-experience and reenactment of the past. Participants reconnected with past online heritage through linguistic adaptation and nostalgic engagement, naturally adopting past styles. Moreover, AI-generated conversational systems offer an innovative methodological tool for investigating historical online behaviors, enabling real-time observation of user engagement with legacy digital contexts. Interactions with RetroChat revealed that past online experiences can evoke online life memories and extend beyond digital spaces into real-life recollections. This suggests that interactive, AI-driven approaches can facilitate deeper engagement with digital heritage, triggering memory flashbacks and personal reflections. Our approach aligns with broader discussions on digital heritage and cultural memory, reinforcing the idea that digital preservation should not merely archive content but also create meaningful opportunities for users to engage with historical digital environments.